\documentclass[final,5p,times,twocolumn]{elsarticle}

\usepackage{amssymb}
\usepackage{amsmath}

\usepackage{natbib}
\usepackage{caption}
\usepackage{subcaption}
\usepackage{xcolor}
\usepackage[margin=10pt,font=small,labelfont=bf,labelsep=endash]{caption}

\journal{Fusion Engineering and Design}

\begin{document}

\begin{frontmatter}



\title{Strategy and guidelines for the calibration of the ITER radial neutron camera}


\author[UU]{M. Cecconello\corref{cor1}}
\ead{marco.cecconello@physics.uu.se}
\author[PL]{R. Miklaszewski}
\author[ENEA]{D. Marocco}
\author[UU]{S. Conroy}
\author[ENEA]{F. Moro}
\author[ENEA]{B. Esposito}
\author[ENEA]{S. Podda}
\author[PL]{B. Bienkowska}
\author[PL]{A. Szydlowski}

\address[UU]{Department of Physics and Astronomy, Uppsala University, SE-751 05, Uppsala, Sweden
}
\address[PL]{Institute of Plasma Physics and Laser Microfusion, Hery Street 23,
	01-497, Warsaw, Poland}
\address[ENEA]{ENEA, C.R. Frascati, Via E. Fermi 45, Frascati, 00044 Rome, Italy}

\cortext[cor1]{Corresponding author. Tel: +46 18 471 3043; fax +46 18 471 3853}

\begin{abstract}
A calibration procedure is proposed for the ITER Radial Neutron Camera. No in-vessel calibration using external neutron sources is required: instead, it is proposed to rely on embedded sources, reference ITER pulses and cross-calibration with ITER fission chambers and activation system coupled to Monte Carlo simulations of radiation transport for the validation of the RNC calibration and for its tracking during ITER lifetime. 
\end{abstract}

\begin{keyword}
ITER \sep Radial Neutron Camera \sep Calibration 


\end{keyword}

\end{frontmatter}



\section{Introduction}
\label{Sec:Intro}
ITER \cite{ITER2007} main goal is to demonstrate the feasibility of delivering energy by means of a burning fusion plasma with a power gain $Q = 10$ via deuterium-tritium (DT) reactions where the necessary tritium fuel is internally bred. 
ITER will be equipped with several diagnostic systems for machine protection, basic and advanced control and physics analysis all aiming at securing its main mission. 
One of such diagnostics is the Radial Neutron Camera (RNC) whose main purpose is the estimation of the neutron emissivity profile $\varepsilon$ and of the fusion power $P_{fus}$ with 10 \% accuracy and precision and a time resolution of 10 ms and a spatial resolution of $a$/10 for $r/a < 0.8$. 
The calibration of the RNC will provide the absolute relationship between the measured counts at the detectors and the corresponding neutron emissivity, that is the relation between the volume integrated neutron emissivity in each field of view and the signal recorded by the corresponding detector. 
As the final accuracy and precision in the $\varepsilon$ and $P_{fus}$ are the result of many factors working in chain, the uncertainty in each detector calibration has to be much better than 10 \%.  
The calibration of NCs is usually done via the \emph{in-situ} method: a neutron source of known strength is located at different positions inside the tokamak while counts are recorded in the detectors. The neutron yield form the source is then coupled to the detectors counts via Monte Carlo simulations. 
Typical neutron sources used for the \emph{in-situ} calibration are $^{252}$Cf for 2.45 MeV neutrons and portable DT Neutron Generators (NGs) for 14 MeV neutrons. All the major tokamaks have been absolutely calibrated in this way: JET \cite{Syme, Batistoni}, TFTR \cite{Jassby} and JT-60 \cite{Nishitani} and a similar approach is considered for ITER NCs \cite{Bertalot}. 
\emph{In-situ} calibration's challenges are the need of very strong neutron sources and long calibration times, the absolute characterization of their yield as well as their handling and positioning inside the vessel. 
The ageing of the detectors and changes in the environment where the diagnostic is located (for example due to the activation of surrounding structural materials) require the RNC calibration to be traced at regular intervals through the entire ITER life time. If the evolution in time of RNC calibration must be monitored via the \emph{in-situ} calibration then access to the vessel must be granted periodically for extended period of times, which  might be an issue on ITER. A review of these and additional issues can be found in \cite{Sadler}.
The aim of this work is to propose a series of guidelines and a strategy for the calibration of the RNC diagnostic over its entire life-time which does not require any \emph{in-situ} calibration.
\section{The radial neutron camera}
\label{Sec:RNC}
The radial neutron camera will be installed on ITER Equatorial Port 1 (EP01) and will measured the neutron flux via an array of collimated lines of sight on a plasma poloidal cross-section. The RNC design has been optimized to provide a uniform coverage of the plasma volume with a Count Rate (CR) $ \leq $ 1 MHz with a 1 \% statistical uncertainty and a 10 ms time resolution from the 17 MA, DT, 700 MW scenario to the 15 MA, DD, 1 MW scenarios \cite{MaroccoIAEA}. 
The RNC consists of: 1) an ``in-port'' sub-system located closest to the plasma, looking at the plasma edge regions with 6 in-port collimators (length from 0.95 to 1.5 m and with 3.2 cm diameter) each equipped with diamonds and $^{238}$U Fission Chambers (FCs) detectors; and 2) an ``ex-port' sub-system between the ``in-port'' sub-system and the bio-shield, whose 16 lines of sight cover the most central region of the plasma (length varying from 1.4 to 3.0 m and with a diameter of 1.1 cm) divided in two fans of 10 lines of sight each each equipped with two scintillators.

\subsection{Characteristics of the RNC detector modules}
\label{Sec:RNCDetChar}
The RNC detectors should be sensitive to 2.5 and 14 MeV neutrons with a selectable energy threshold for background rejection and the minimization of the scattered neutrons contribution. 
They should be able to discriminate between 2.5 and 14 MeV neutrons and between neutrons and $\gamma$-rays. The required dynamic range should be $10^4$ and the detectors should be able to support CRs up to 1 MHz so that a statistical error of 1 \% (10 ms integration time) of 3 \% (1 ms integration time) can be obtained. 
The detector should have an energy resolution of approximately 3 \% at 14.1 MeV and of 7.5 \% at 2.5 MeV in order to allow the measurement of the plasma temperature of about 5 keV. 
In order to provide the required dynamic range, single crystal diamonds (for DD and DT scenarios) and FCs (for high power DT operations) will be used in the ``in-port'' subsystem. Each FC will have a deposit of 2 mg/cm$^2$ of $^{238}$U per layer (9 layers are foreseen per FC).
The ``ex-port'' channels will be equipped with two scintillators (coupled to low magnetic field sensitivity PMTs) of 1 and 10 cm thickness (for high and low power operations respectively) and will be actively cooled.

\subsection{Maximum uncertainty in the detection efficiency}
\label{Sec:CalErr}
The requirement to reconstruct $\varepsilon$ and $P_{fus}$ with an accuracy and precision $<$ 10 \% imposes a maximum uncertainty in the measured detector CRs. Uncertainties in CRs are due to uncertainties in the detector $\epsilon$, the sight-line solid angle $\Omega$ and to the contribution from scattered neutrons. 
Typical uncertainties in scintillators, FCs and diamonds $\epsilon$ are in the range 1 - 5 \%, 10 \% and 5 \% respectively.
The main sources of uncertainties in the scintillator $\epsilon$ are the energy calibration, the neutron and $ \gamma $-rays light output functions and the (n,C) cross-sections for $ E_n > 8 $ MeV. 
Sufficiently high counting statistics allows the scintillator calibration to be determined with less than 1 \% relative error. Typical uncertainties in the light output at $ E_n = $ 2.45 MeV and 14 MeV are approximately 4 \% resulting in $\delta \epsilon / \epsilon \approx$ 5 \% while uncertainties in the energy resolution function have a negligible effect. The uncertainty in $ \Omega $ has been estimated as $\delta \Omega = 2 k [4 (\delta D/D)^2 + (\delta \ell/\ell)^2]^{1/2} $
where $\Omega(D,\ell) = \pi D^4/64 \ell^2$, $\delta D = 0.1$ mm and $\delta \ell = 1$ mm giving $ \delta \Omega/\Omega \approx 0.04 $.
The signal (direct neutrons) to noise (scattered neutrons) ratio has been calculated with MCNP to be $ \approx $ 33 \cite{MORO20171033}. 
The combined effect of all these uncertainties and of poissonian counting statistic on $\varepsilon$ has been estimated resulting in an accuracy and precision of the reconstructed profile $<$ 10 \% for $r/a < 0.8$. Since $ \Omega/\Omega \approx 0.04 $, the $\delta\epsilon/\epsilon$ must not exceed 5 \%.  
\section{RNC calibration}
\label{Sec:Cal}
\subsection{Pre-delivery calibration}
\label{Sec:CalPre}
The pre-delivery phase, performed at the EU Domestic Agency (DA) facility,  will measure the detectors’ response functions, i.e. the pulse height spectrum corresponding to mono-energetic neutrons and $\gamma$-rays from well characterized radiation sources. 
The response function provides the detector $\epsilon$ in terms of counts per incident neutron above a given energy threshold and, for diamonds and scintillators, the detector pulse height energy resolution. The availability of the detector’s response function in a wide $ E_n $ range enables the use of diamond and scintillator as spectrometers and provides, in the case of scintillators, the relation between the energy of the recoil proton and the scintillation light output and, in the case of diamonds, the relation between the neutron energy and deposited energy in the the detector. Detector power supply voltages and data acquisition settings should be clearly stated in the pre-delivery calibration and used at each following calibration. 

\subsubsection{Scintillators characterization}
\label{Sec:CalPre_Scint}
The response function for $ E_n = $ 2.45 and 14 MeV will be measured for all scintillators using DD and DT neutron generators providing $\epsilon(E_n)$ and the energy resolution. The maximum uncertainty in the light output function should not exceed 5 \% for $ E_n \in [2, 20]$ MeV. 
Due to budget limitations, only a sub-set of scintillators will be fully characterized by measuring the response functions and the recoil proton light output function in a wide range of neutron energies (2 – 20 MeV) using a neutron white source with the associated-time of flight technique. The response function will be used to provide $\epsilon$ as a function of the acquisition energy threshold.
The response function for the remaining scintillators will be obtained by cross-comparison of the DD and DT neutron spectra with the fully characterized detectors. 
The scintillators will be also calibrated using $\gamma$-ray sources ($^{22}$Na, $^{137}$Cs and $^{207}$Bi) to determine the recoil electrons light output function. 
The electron equivalent energy scale of all detectors will be determined with an uncertainty less than 1 \% using Compton edge and forward fitting methods by acquisition of pulse height spectra with a number of counts sufficiently high at the relevant energies to make the counting statistical uncertainty negligible.

\subsubsection{Diamonds characterization}
\label{Sec:CalPre_Diam}
The energy scale of all diamond detectors will be measured using $^{239}$Pu, $^{241}$Am, $^{244}$Cm $\alpha$ particles sources with an uncertainty less than 1 \% at three well separated energies ($\approx$ 5.2 MeV, 5.5 MeV and 5.8 MeV).
Response functions to 2.45 and 14 MeV neutrons will be measured for all diamonds using DD and DT neutron generators.
A sub-set of the diamonds will be fully characterized by measuring their response function in a wide range of neutron energies (1 – 20 MeV) and at 2.45 and 14 MeV providing $\epsilon$, the pulse height resolution function and the deposited energy as a function of the neutron energy. The required uncertainty in the counting statistics in the $^{12}$C(n,$\alpha$)$^9$Be peak area should not exceed 2 \% and 5\% in the $\epsilon$.
Considering the present day technology, extension of the results of the fully characterized detectors to the remaining detectors might be problematic because of the irreproducibility in the manufacturing of the electrodes contacts.

\subsubsection{Fission Chambers characterization}
\label{Sec:CalPre_FC}
Efficiency calibration for all FCs must be carried out because of the inherent variability of the thickness of the deposited $^{238}$U layer which will determined at an accelerator facility using 2.5 and 14 MeV neutrons. In addition, functionality checks and determination of optimal $ \alpha $-particle discrimination threshold settings using $^{252}$Cf sources will be carried out.

\subsubsection{Pre-delivery calibration of the assembled RNC}
\label{Sec:CalPre_RNC}
It is foreseen that the RNC will be assembled once at the DA during the pre-delivery phase. A photogrammetry survey will be carried out to determine the position and the alignment of the detector-collimator system with respect to external reference points on the Diagnostic Shielding Module (DMS) for both the ``in-port'' and ``ex-port'' systems. This survey will be used to determine the solid angle of the sightlines and to validate analytical calculations of $ \Omega $.
These will be further validated by the measurement of the neutron fluxes at the detectors by using neutron sources placed at the front end of the collimators. The relative uncertainty $ \delta\Omega / \Omega $ must not exceed 5 \%. The photogrammetry survey will be used in the post-delivery phase to verify the conformity of the assembling and installation of the RNC at ITER. 
The neutron cross-talk between detectors in adjacent lines of sight will also be measured in this phase.
Finally, reference pulse height spectra will be measured using embedded radio-active sources for all the detectors and simulated via MCNP using the response functions determined as described in sections 3.1.1 to 3.1.3.

\subsection{Post-delivery calibration and in-vessel neutron calibration}
\label{Sec:CalPost}
Prior to the installation of the ``in-port'' and ``ex-port'' subsystems on EP01, functional tests of the detectors will be performed using the embedded sources to assess any deviation in the detectors’ performance from those measured at the DA. These measurements will be repeated after the installation of the RNC on ITER.
The alignment of the collimator-detector geometry for the ``in-port'' and ``ex-port'' subsystems will be checked both prior and after their installation on ITER by photogrammetry to relate on an absolute reference system the position of the detectors and collimators with respect to the vacuum vessel and additional reference points (to be determined). 
ITER is planning an in-vessel calibration for all the neutron diagnostics to be carried out with a $^{252}$Cf source and portable DD and DT generators \cite{Bertalot}. This in-vessel calibration, supported by MCNP calculations, might be useful to check the alignment and absolute positions of RNC components. However, the accuracy in the positioning of the neutron source in front of the RNC sight-lines is of the same order of the RNC collimator diameters making accurate alignment challenging. Comparison of the in-vessel calibration measurements with MCNP calculated fluxes at detector positions in order to check the RNC alignment is therefore not considered useful. Acceptable tolerances in the RNC positions are 5 mm.

\subsection{Routine calibration}
\label{Sec:CalRout}
The routine calibration consists in the regular, periodic monitoring of the detector calibration (threshold and response function) throughout ITER life time. This will be achieved by a combination of sets of embedded radioactive sources installed near the detectors, of reference ITER plasma discharges and of cross-calibrations with other neutron diagnostics (namely FCs and activation systems) all supported by MCNP simulations.
During ITER life-time, the properties of the RNC detectors may change due to short term environmental variations (such as changes in the stray magnetic field, CRs and temperature during a plasma discharge) and to long term drifts caused by the ageing and deterioration of the detectors due to radiation damage (particularly in the case of scintillators and diamonds) and to an increased activation background. One-off events, such as abnormal deviations (earthquakes, failure of one or more of the RNC support systems such as detector cooling) might also impact the RNC calibration. 
The calibration proposed here assumes that both the pre- and post-delivery characterizations have been carried out at a well defined temperature and that subsequent calibrations must be carried out at the same temperature. If this is not possible, that a method of tracking the temperature dependence of the calibration should be established. 

\subsubsection{Short-term routine calibration: in-shot calibration}
\label{Sec:CalInShot}
Scintillators are known to gain changes when exposed to high CRs ($>$ 0.5 MHz), strong B fields or temperature variations. Monitoring of the relative gain shift during an ITER pulse will be done using an external LED whose amplitude is sufficiently stable for the duration of an ITER pulse and by monitoring the location, in the recoil proton pulse height spectrum, of the ``head-on'' collision edge for DD and DT neutrons. Correction for gain variations, thanks to the full digital data acquisition system (FPGAs), can be done during an ITER pulse with a duty cycle shorter than the time resolution requirement for quantities needed for basic control (for example $P_{fus}$). 
Long-term variations of the LED system (source, fiber optics and light wave-guides) due to irradiation do not affect the in-shot calibration. Such long-term changes will be monitored thanks to the embedded $\gamma$-rays sources.
FC and diamond detectors are not expected to be affected by such effects and therefore no in-shot calibration is required.

\subsubsection{Medium-term routine calibration: inter-shot procedure}
\label{Sec:CalInterShot}
To ensure the stability of the detector system in terms of energy scale calibration and pulse height resolution, routine measurements with embedded $\gamma$-rays, $\alpha$-particles and neutron sources should be made regularly, possibly at the end of each ITER plasma discharge. Analysis of the PHSs collected during such inter-shot calibration will provide information on the actual pulse height resolution and will allow to follow their changes in time. 

\subsubsection{Long-term routine calibration}
\label{Sec:CalLongTerm}
The stability of the detectors' neutron $\epsilon$ at 2.5 MeV and 14 MeV should be checked two or three times a year for the entire ITER life-time or, in case of severe deviations detected during short- and medium-term calibrations, as soon as possible. This long term calibration requires a well defined neutron source with a known strength and energy spectrum. 
The use of a DD/DT neutron generator inside the RNC itself has been considered and discarded. The main advantage provided by such a system would be the capability of calibrating all the RNC detectors in approximately 4 hours (assuming a DT yield of 10$^8$ ns$^{-1}$) with a 1 \% statistical error on the number of counts and with a high energy resolution (Full Width Half Maximum $<$ 100 keV). However, engineering integration constraints (no space available for the ``in-port'' subsystem and reduction of internal shielding and collision with the high-resolution neutron spectrometer collimator for the ``ex-port'' subsystem) and difficulties of tracking the absolute yield and anisotropy of the generator itself resulted in the rejection of this approach.  
Instead dedicated reference DD and DT plasma discharges, dominated by the thermal component, should be used for this similarly to what has been successfully employed at JET and TFTR. Presence of a strong thermal emission component is typical of the majority of the foreseen DT scenarios (DT ohmic plasmas of at least 5 MW $P_{fus}$ are deemed sufficient for the neutron $\epsilon$ determination) while for DD, dedicated Ohmic discharges will be needed. 
These reference plasma discharges should preferably have a low $ T_i $ in order for the neutron energy spectrum to have a small FWHM ($\propto \sqrt{T_i}$). A simultaneous cross-check of the RNC with FCs and activation measurements will provide the source strength. 
Estimated calibration time for all channels is $ \approx $ 100 s to ensure a 1 \% statistical error on the number of counts.

\subsection{Embedded calibration sources}
\label{Sec:CalEmbeddedSources}
The ``In-Port'' and ``Ex-Port'' detectors will be equipped with embedded radiation sources to be used in all calibration phases. The calibration sources must have half-lives comparable to ITER life time ($\approx$ 20 y) and an activity sufficient to provide calibration counts 10 times larger than background counts. For diamonds and scintillators, a counting statistics error $ \leq$ 1\% should be achievable during the inter-shot period ($\approx$ 1000 s). For FCs, $10^4$ counts should be accumulated in the whole neutron pulse height spectrum for $\alpha$-particles discrimination threshold.
For scintillators $^{207}$Bi $\gamma$-ray sources (570 keV, 1064 keV and 1770 keV with $t_{1/2} \approx 31.55$ y) will be used. The required activity per source has been estimated to be 0.2 MBq giving $10^6$ counts in 1000 s. Taking into account the relative intensities of the three main $ \gamma $-rays energies and the fraction of counts contributing to the corresponding measured Compton edges, the statistical error can be limited at $\leq$ 1 \%.
For diamonds, $\alpha$-particle source consisting of mixed nuclides $^{239}$Pu, $^{241}$Am, $^{244}$Cm ($E_\alpha$ = 5.2 MeV, 5.5 MeV and 5.8 MeV with $t_{1/2} \approx$ $2.4 \times 10^4$, 432 and 18 y respectively). Diamonds are insensitive to $ \gamma $-rays (including those from activation) and therefore the required activity has been estimated to 6 kBq per source corresponding to a CR of 1.5 kHz resulting in an achievable energy calibration relative uncertainty of approximately 0.14 \%.
A suitable calibration source for FCs is $^{252}$Cf for which $ t_{1/2} \approx $ 2.6 y. The contribution from $ \gamma $-rays to the calibration of FCs does not represent a problem because of the smaller energy released per interaction compared to the fission products. The estimated required activity is $\approx$ 5 MBq per source corresponding to a deposited mass of 0.26 $\mu$g of $^{252}$Cf given a CR of $ \approx $ 1 Hz which is sufficient since it is intended only for the long-term routine calibration.
The contribution of such $^{252}$Cf source on diamond detectors, assuming an efficiency of $ \approx $ 4 \% at fission spectrum energies and a distance to of $\approx$ 20 cm, has been estimated to 50 Hz, which can be however removed thanks to the diamonds $ \alpha $/n pulse shape discrimination capability.
However, a 5 MBq $^{252}$Cf would result in a contact dose of $\approx$ 0.1 Sv/h while the hands-on limit is 100 $\mu$Sv/h. Two are the alternatives considered: \emph{i}) an external source to be used during the post-delivery and/or \emph{ii}) ITER reference plasma discharges as discussed in section \ref{Sec:CalLongTerm}. With an efficiency of $ \approx 0.5 \times 10^{-5}$ counts per neutron, $ \approx 100 $ s would be sufficient for a calibration with good statistics.

\section{Conclusions}
\label{Sec:Conclusions}
The proposed RNC calibration relies on the characterization of the detectors before and after its installation on ITER. Embedded radiation sources together with reference plasma discharges, cross-calibration with ITER FCs and activation systems coupled to MCNP radiation transport calculations are sufficient to monitor the RNC calibration for the entire ITER lifetime. No in vessel calibration of the RNC is required.
The procedures here described will necessarily have to be revised and updated to take into account the on-going development of the design of the RNC. In particular, once the geometry and material composition of the RNC is finalized, a proper assessment of the $\gamma$-ray background at the detectors' location will need to be carried out to determined the intensity of the embedded radio-active sources. 
The alignment of the detector-collimator-plasma and knowledge of the relative positions of the detectors and collimators with respect to the vacuum vessel is a crucial factor for the proper operation of the RNC. Photogrammetry is proposed both in pre-delivery and post-delivery phase, but a dedicated study is necessary. Remote handling and health and safety considerations associated with the RNC calibration have not been addressed in this work but should be in the future.

\section*{Acknowledgements}
The work leading to this publication has been funded partially by Fusion for Energy under the Contract F4E-FPA-327. This publication reflects the views only of the author, and Fusion for Energy cannot be held responsible for any use which may be made of the information contained therein. The views and opinions expressed herein do not necessarily reflect those of the ITER Organization.

\bibliography{RNC_Calibration_Paper}
\end{document}